\documentclass[12pt]{article}

\usepackage[totalheight = 23cm, totalwidth = 17cm]{geometry}
\usepackage{amssymb,amsmath,amsfonts,amsbsy,graphicx,psfrag,cite,caption}

\newcommand{\mpl}{m_{\rm Pl}}

\newcommand{\calH}{{\cal H}}
\newcommand{\calL}{{\cal L}}
\newcommand{\calO}{{\cal O}}
\newcommand{\calP}{{\cal P}}
\newcommand{\calR}{{\cal R}}

\newcommand{\calI}{{\cal I}}
\newcommand{\calC}{{\cal C}}

\newcommand\be{\begin{equation}}
\newcommand\ee{\end{equation}}
\newcommand\ba{\begin{eqnarray}}
\newcommand\ea{\end{eqnarray}}\newcommand\eq{\begin{equation}}           
\newcommand\en{\end{equation}}

\begin{document}

\begin{titlepage}

%\rightline{\footnotesize{CTPU-14-XX, APCTP-Pre2014-015, other preprint number}}
\rightline{\footnotesize{CTPU-14-13,APCTP-Pre2014-015}}

\begin{center}

\vskip 1.0cm

{\Huge CMB probes on the correlated axion isocurvature perturbation}

\vskip 1.0cm

\large{
Kenji Kadota$^a$,
\hspace{0.1cm}
Jinn-Ouk Gong$^{b,c}$,
\\
Kiyotomo Ichiki$^{d,e}$
\hspace{0.1cm} and \hspace{0.1cm}
Takahiko Matsubara$^{d,e}$
}

\vskip 0.5cm

\small{\it 
$^{a}$Center for Theoretical Physics of the Universe, Institute for Basic Science, Daejeon 305-811, Korea
\\
$^{b}$Asia Pacific Center for Theoretical Physics, Pohang 790-784, Korea 
\\
$^{c}$Department of Physics, Postech, Pohang 790-784, Korea
\\
$^d$Department of Physics, Nagoya University, Nagoya 464-8602, Japan
\\
$^e$Kobayashi-Maskawa Institute for the Origin of Particles and the Universe
\\
Nagoya University, Nagoya 464-8602, Japan
}

\vskip 1.2cm

\end{center}

\begin{abstract}

We explore the possible cosmological consequence of the gravitational coupling between the inflaton and axion-like fields. In view of the forthcoming cosmic microwave background (CMB) polarization and lensing data, we study the sensitivity of the CMB data on the cross-correlation between the curvature and axion isocurvature perturbations. Through a concrete example, we illustrate the explicit dependence of the scale dependent cross-correlation power spectrum on the axion parameters.

\end{abstract}

\end{titlepage}

\setcounter{page}{0}
\newpage
\setcounter{page}{1}

\section{Introduction}

The ever-growing precision of the CMB measurements can offer an excellent link between the fundamental physics predictions and the observational data. While the current CMB data are consistent with the simple $\Lambda$CDM with pure adiabatic perturbation, there is till room for seeking a model beyond such a simple parameterization which can naturally arise in the particle theoretical studies of the early Universe \cite{plpara,pie,enq,buc2,crott,sil}. The light species such as the neutrinos and axion-like light particles, as well as the inflation models beyond the simple single field inflation, for instance, have been the subjects of studies in view of the forthcoming cosmological data \cite{kev4,les,wan,sai,jou,bir,bat,takeu,david3,rin,pedro,davi,gra,kamao,hlo,pol,lang,bri,kadotawayne,les2,mart2,mart}. We in this article are particularly interested in the isocurvature perturbation by axion-like fields which can correlate with the curvature perturbation through gravitational couplings. We explicitly show that the coupling between the inflaton and the axion-like fields can lead to a non-trivial scale dependent cross-correlation between the curvature and isocurvature perturbations. There can indeed exist ubiquitous light degrees of freedom in the early Universe, such as pseudo-Goldstone bosons (PGB) from spontaneous symmetry breaking of some approximate symmetries, and the isocurvature perturbation is expected to generically arise besides the dominant adiabatic perturbation \cite{cro,aaron,bel,kur,mack,arv,acha,li,bar,fer}. We just refer to these light PGB as the axion in this article, and our discussions are straightforwardly applicable to any axion-like light fields.

To motivate the cosmological study of the cross-correlated isocurvature perturbation, Section~\ref{fore} presents the Fisher likelihood analysis to clarify the sensitivity of the forthcoming CMB ($E$-mode) polarization and CMB lensing data by a Planck-like satellite experiment on the cross-correlation power spectrum. We then derive an analytic expression for the cross-correlation power spectrum between the curvature and isocurvature perturbations, and obtain the upper bound on the cross-correlation power spectrum amplitude in terms of the axion parameters in Section~\ref{example}.

\section{Forecasts}
\label{fore}

The aim of this section is to see how much the cosmological parameter estimations can be affected by the cross-correlated isocurvature perturbation, which would serve to motivate our study of the axion isocurvature cross-correlation power spectrum in the next section.  
For forecasting the bounds on the cosmological parameter estimations using the forthcoming CMB data, we perform the Fisher matrix analysis outlined below. 
We consider a Planck-like CMB satellite experiment and the CMB observables $(T,E,L)$ of our interest are, respectively, the CMB temperature, $E$-mode polarization and the CMB deflection angle representing the CMB lensing \cite{lewis}. 
The Fisher matrix is then given by \cite{kar}
%% \ba
%% F_{ij} = \sum _{\ell} \frac{2\ell+1}{2} f_\text{sky} \text{Tr} \left[{\bf \tilde C }_\ell^{-1}(\vec{p})  \frac{\partial {\boldsymbol C}_\ell}{\partial p_i} (\vec{p}) {\bf \tilde C}_\ell^{-1}(\vec{p})  \frac{\partial{{\boldsymbol C}_\ell}}{{\partial p_j}}(\vec{p}) \right]
%% \ea
%% %

\ba
F_{ij} = \sum _{\ell} \frac{2\ell+1}{2} f_\text{sky} \text{Tr} \left[  \frac{\partial {\boldsymbol C}_\ell}{\partial p_i} (\vec{p}) {\bf \tilde C}_\ell^{-1}(\vec{p})  \frac{\partial{{\boldsymbol C}_\ell}}{{\partial p_j}}(\vec{p}) {\bf \tilde C }_\ell^{-1}(\vec{p})
\right]
\ea
with % ${\boldsymbol C}_\ell$,
\begin{equation}
{\bf \tilde C}_\ell =
{\boldsymbol C}_\ell+{\boldsymbol N}_\ell
=
 \left( \begin{array}{ccc}
C_\ell^{TT}+N_\ell^{TT} & C_\ell^{TE} & C_\ell^{TL} \\
C_\ell^{TE}& C_\ell^{EE}+N_\ell^{EE} & 0  \\
C_\ell^{TL} & 0  & C_\ell^{LL}+ N_\ell^{LL} \\
\end{array} \right) \, ,
\end{equation}
where $C_\ell^{XY}$ and $N_\ell^{XY}$ are respectively the power spectra of the CMB signal and the noise in the measurements. $\vec{p}$ is the vector consisting of the cosmological parameters $\{p_i\}$. We assume for concreteness the Planck-like experiment covering up to the multipole of $l_\text{max}=2500$, the sky coverage of $f_\text{sky}=0.65$ and three frequency channels 100, 142 and 217 GHz, where the beam width $\theta_\text{FWHM}$ [arcmin] and the temperature (polarization) sensitivity $\Delta^T(\Delta^P)$ [$\mu K$/pixel] are, respectively, $(\theta_\text{FWHM},\Delta^T,\Delta^P)=(9.5',6.8,10.9)$, $(7.1',6.0,11.4)$ and $(5.0',13.1,26.7)$. For the temperature and polarization noise, we simply consider the dominant detector noise represented by the photon shot noise \cite{knox95,bond97}, and, for the statistical noise of the CMB lensing deflection field, we use the optimal quadratic estimator of Hu and Okamoto \cite{take,oka2}. To study the isocurvature perturbation, we introduce $(A_\calI, A_\calC,n_\calI)$ in addition to the conventional six $\Lambda$CDM parameters ($\Omega_\Lambda=0.69$, $\Omega_m h^2=0.14$, $ \Omega_b h^2=0.022$, $n_\calR=0.96$, $A_\calR=2.2\times 10^{-9}$, $\tau$ (reionization optical depth)$=0.095$) with the numerical values being the fiducial values in our Fisher analysis \cite{plpara}. The spectral index of the cross-correlation is set to $n_\calC=(n_\calR+n_\calI)/2$ for simplicity, which is indeed realized in and motivated from our axion model in the next section.
The total matter density consists of baryon and (non-baryonic) cold dark matter (CDM) $\Omega_m=\Omega_b+\Omega_{c} = 1-\Omega_\Lambda$. We assume the flat Universe and use the reduced Hubble parameter $h=\sqrt{\Omega_m h^2/(1-\Omega_{\Lambda})}$ in our analysis. We define the power spectra of the curvature, isocurvature and their cross-correlation, denoted by subscripts $\calR$, $\calI$ and $\calC$ respectively, as
\begin{equation}
\calP_X = A_X(k_0) \left( \frac{k}{k_0} \right)^{n_X-1} \, ,
\end{equation}
with $X \supset (\calR,\calI,\calC)$, and the fractions of the isocurvature perturbation and cross-correlation as
\begin{align}
\beta_\calI  = \frac{\calP_\calI}{\calP_\calR} \, ,~
\beta_\calC  = \frac{\calP_\calC}{\sqrt{\calP_\calR\calP_\calI}} \, ,
\end{align}
respectively.
%\ba
%\calP_{\calR}=\frac{k^3}{2\pi^2}P_{\calR\calR}=A_S (k_0)\left(\frac{k}{k_0}\right)^{n_s-1}
%\ea
%\ba
%\calP_{\calI}=\frac{k^3}{2\pi^2}P_{\calI\calI}=A_I (k_0)\left(\frac{k}{k_0}\right)^{n_{iso}-1}
%\ea
%\ba
%\calP_{\calC}=\frac{k^3}{2\pi^2}P_{\calR\calI}=A_C (k_0)\left(\frac{k}{k_0}\right)^{n_{c}-1}
%\ea
%\ba
%\beta_{iso}=\frac{\calP_{\calI}}{\calP_{\calR}},~~\beta_c=\frac{\calP_{\calC}}{\sqrt{\calP_{\calR} \calP_{\calI}}}
%\ea
Unless stated otherwise, $A$'s and $\beta$'s are evaluated at the reference scale $k_0=0.05$ Mpc$^{-1}$ and the isocurvature fraction is set to $\beta_\calI=0.04$ (95\% CL upper bound from Planck+WMAP \cite{plinf}) in the following analysis. We modified the CAMB \cite{camb} to calculate the CMB power spectra in existence of the isocurvature cross-correlation for our purpose. We found the sign of $\beta_\calC$ did not affect our conclusion quantitatively, and the following discussions simply assume a positive $\beta_\calC$.  
The Fisher matrix consists of aforementioned 9 parameters, and the marginalized errors for the parameters involving the isocurvature perturbation are listed in Table \ref{table:error} for different cross-correlation power spectrum amplitudes. %The constraints on $A_\calI$ and $n_\calI$ are not expected to change significantly for $\beta_\calC \lesssim \calO (0.01)$ which is sufficiently small compared with the isocurvature perturbation whose amplitude is fixed to $\beta_\calI=0.04$ in our analysis.

\begin{table}[htb!] 
\begin{center}
\begin{tabular}{|c||c|c|c|c|}
\hline
  & $T$ & $TE$ & $TL$ &Joint \\
\hline\hline

$\beta_\calC=1$ &    & & & \\
 $\sigma (n_\calI)/ n_\calI$ &  33  &    13 &  21 &    12 \\
 $\sigma (A_\calI)/ A_\calI$   &  240 &     81  &    220  &   80  \\
$\sigma(A_\calC)/ A_\calC$   &  65 &      11 &     20 &       11 \\ 

 \hline
 $\beta_\calC=0.1$ &  &    & & \\
 $\sigma (n_\calI)/ n_\calI$ &    110  &    39  &  65   &    38  \\
 $\sigma (A_\calI)/ A_\calI$   &    260  &   100      &  260  &   100 \\
$\sigma(A_\calC)/ A_\calC$  &       230 &    76 &   170  &           74\\ 
 \hline
 $\beta_\calC=0.01$ &  &  &  & \\
 $\sigma (n_\calI)/ n_\calI$ &   150   &  49   &   85   &    47 \\
 $\sigma (A_\calI)/ A_\calI$ & 290    &  110   &   280 &    110   \\
$\sigma(A_\calC)/ A_\calC$ &    1800  & 710 & 1700 &        690 \\ 
%%  $\beta_\calC=1$ &    & & & \\
%%  $\sigma(A_\calC)/ A_\calC$ & 52 &   3.8 &     7.6 &     3.8 \\
%%  $\sigma (A_\calI)/ A_\calI$ & 52 &   12  &   30  & 12 \\
%%  $\sigma (n_\calI)/ n_\calI$ & 11 &   4.3  & 6.7  &  4.2 \\
%% \hline
%%  $\beta_\calC=0.1$ &  &    & & \\
%%  $\sigma(A_\calC)/ A_\calC$ &       98  &       24 &     55 &     23 \\
%%  $\sigma (A_\calI)/ A_\calI$ &   67   & 13        & 31 &  13 \\
%%  $\sigma (n_\calI)/ n_\calI$ &   17 &  6.1   & 9.9  &  6.0 \\
%% \hline
%%  $\beta_\calC=0.01$ &  &  &  & \\
%%  $\sigma(A_\calC)/ A_\calC$ &   600 &   230 &   540 &   220 \\
%%  $\sigma (A_\calI)/ A_\calI$ &  70   &   13    &   32  &  13  \\
%%  $\sigma (n_\calI)/ n_\calI$ &  18   &  6.4  & 11 &   6.3 \\
 \hline
\end{tabular}
\end{center}
\caption{1$\sigma$ errors [\%] for different values of $\beta_\calC$. $T$ refers to the analysis using only the CMB temperature data. $TE$ ($TL$) refers to the analysis using both temperature and polarization (temperature and lensing) information. Joint refers to the use of $T$, $E$ and $L$.}\label{table:error}
\end{table}

\begin{figure}[htb!]
\begin{center}    
      \psfrag{l}[B][B][1][0]{$\ell$}
      \psfrag{kenji1234567891234}[B][B][1][0]{$\ell(\ell+1)\partial C_\ell/\partial \ln A_\calC /2\pi$}
     \includegraphics[width= 0.48\textwidth]{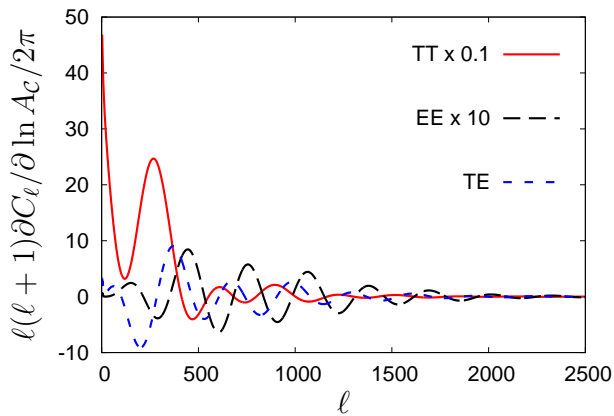}
% produced by nov6.gnu
\end{center}
\caption{The derivatives $C_\ell^{XY}(XY=TT,EE,TE)$ with respect to $A_\calC$ (for $\beta_{\calI}=0.04,\beta_\calC=1$). For an easier comparison, we plot scaled values for $TT$ and $EE$ by a factor 10 as indicated in the figure.}
 \label{DelC}
   \end{figure}

\begin{figure}[htb!]
\begin{center}    
      \psfrag{kenji1}[B][B][1][0]{$\tau$}
      \psfrag{kenji2}[B][B][1][0]{$A_\calC (\times 10^9)$}
     \includegraphics[width= 0.48\textwidth]{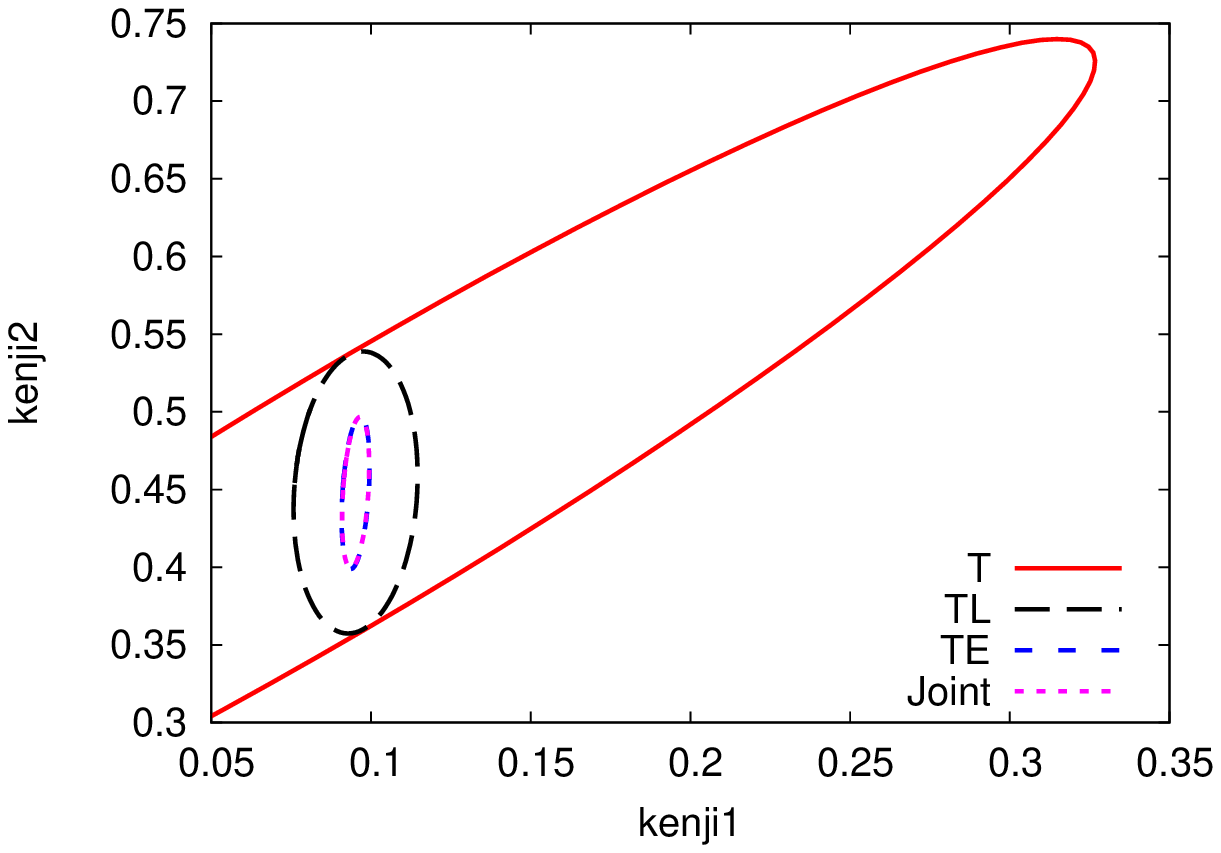}
\end{center}  
\caption{Marginalized $1\sigma$ error contours for $\tau$ and $A_\calC$ (magnified by $10^9$) (for $\beta_{\calI}=0.04,\beta_\calC=1$).}        
\label{tauvsac}
\end{figure}

Let us take a look at the bound on $A_\calC$ which is of our particular interest. We can see that $A_\calC$ is constrained more tightly for a bigger $\beta_\calC$, which is reasonable because a bigger value of $\beta_\calC$ can let the cross correlation make a bigger contribution to the observable total power spectrum. We find that $\beta_\calC \gtrsim \calO(0.1)$ is required for the error of $A_\calC$ as well as those of $A_\calI$ and $n_\calI$ not to exceed 100\%. We can also find an improvement in $\sigma(A_\calC)/A_\calC$ by adding the CMB polarization (and to a lesser extent by the CMB lensing). An advantage of adding the polarization data to the temperature data is that the sensitivity of polarization to $A_\calC$ is different from that of temperature. This is illustrated in Figure \ref{DelC} where we plot $\partial C_\ell/\partial \ln A_\calC$ for the temperature and polarization data. Another, more important reason for the improvement comes from the breaking of the degeneracy among the cosmological parameters. We here particularly point out the degeneracy between $\tau$ and $A_\calC$ which arises because the power spectrum amplitude is suppressed by the reionization optical depth by a factor $\sim e^{-2\tau}$.
% and constraining $\tau$ from the polarization which is sensitive to the reionization bump at $\ell \lesssim 10$ can improve the constrains on $A_\calC$. 
This is illustrated in Figure \ref{tauvsac}. We can clearly see the big degeneracy between $\tau$ and $A_\calC$ in the temperature data alone, which is broken by adding the polarization data. Polarization is sensitive to the reionization bump on large scales ($\ell \lesssim 10$) which can lift the degeneracies concerning $\tau$, resulting in the improved constraints on $A_\calC$. The CMB lensing also improves the constraints on $A_\calC$ because the lensing is sensitive to the initial power amplitude even though polarization would be more powerful in constraining $A_\calC$,  assuming the noise and angular scale of a Planck-like CMB experiment.

%% %%%%%%%%%%%%%%%%%%%%%%%%%%%%%%%%%
%% \begin{figure}[htb!]
%% \begin{center}    
%% \epsfxsize = 0.48\textwidth
%% \epsffile{./CelllnAc1.eps}
%% \end{center}  
%% \caption{ $\partial C_l/\partial \ln A_{\calC} l (l+1)/2\pi$. \kenji{Why is EE zero for low ell? Maybe because of a small amplitude?}}        
%% % figure produced by nov6.gnu
%% %pwd
%% %/Users/kenji/SkyDrive/work/code/isocurvatureIchiki/ichikiiso/cmblensingoutput/nov6.gnu
%% \label{DelC}
%% %\end{figure}

%% %%%%%%%%%%%%%%%%%%%%%%%%%%%%%%%%%
%% \begin{figure}[htb!]
%% \begin{center}    
%% \epsfxsize = 0.48\textwidth
%% \epsffile{./CelllnAc2.eps}
%% \end{center}  
%% \caption{The derivatives with respecto to $A_C$. For an easier conparison, we plotted the scaled values $0.1\times \partial C^{TT}_l/\partial \ln A_C$ and $10 \times \partial C^{EE}_l/\partial \ln A_C$.}
%% % $\partial C_l/\partial \ln A_{\calC} l (l+1)/2\pi$.}        
%% % figure produced by nov6.gnu
%% %pwd
%% %/Users/kenji/SkyDrive/work/code/isocurvatureIchiki/ichikiiso/cmblensingoutput/nov6.gnu
%% \label{DelC}
%% \end{figure}

%%%%%%%%%%%%%%%%%%%%%%%%%%%%%%%%%

\begin{table}[htb!]
\begin{center}
%\caption*{Normalized error $\sigma/\sigma_\text{no iso}$} 
\begin{tabular}{|c||c|c|c|c|c|c|}
\hline
 $$ &  $\Omega_{\Lambda}$ &  $\Omega_m h^2$  &  $ \Omega_b h^2$   &  $n_\calR$ &  $A_\calR$  & $\tau$  \\
\hline\hline
% $\beta_\calC= 1$  & 1.1 & 1.1 & 1.0  &1.4  &  0.82 &  0.79 \\
% $\beta_\calC=0.01$  & 1.1  &1.1  & 1.0  &1.4  & 0.95 & 0.93  \\
%  No correlation  & 1.0 &  1.0 & 1.0 & 1.1 & 0.95 &0.93 \\
 $\beta_\calC= 1$  & 1.1 & 1.1 & 1.0  &1.4  &  0.97 &  0.94 \\
 $\beta_\calC= 0.1$  & 1.1 & 1.1 & 1.0  &1.4  &  1.1 &  1.1 \\
% $\beta_\calC=0.01$  & 1.1  &1.1  & 1.0  &1.4  & 1.1 & 1.1  \\
  No correlation  & 1.0 &  1.0 & 1.0 & 1.1 & 1.1 &1.1 \\
\hline
\end{tabular}
\caption*{Normalized error $\sigma/\sigma_\text{no iso}$} 
\end{center}
\caption{The comparison between the error estimation assuming the isocurvature perturbation ($\beta_\calI=0.04$) and that assuming the $\Lambda$CDM with no isocurvature perturbation. All the errors here are estimated using all of $T$, $E$ and $L$.}
\label{normerror}
\end{table}
The cosmological parameters are in fact not totally independent from each other, and the existence of a small cross-correlation power spectrum can still affect the other cosmological parameters which are well constrained by the CMB alone. 
This is illustrated in Table \ref{normerror} where the errors estimated assuming the isocurvature perturbation are normalized to those assuming no isocurvature components. The marginalized errors in this table are calculated by using the $9\times9$ Fisher matrix except the last row with no cross-correlation which used $8\times8$ Fisher matrix without $A_\calC$. These errors are then divided by those calculated by $6\times6$ Fisher matrix in the $\Lambda$CDM. The error in $\tau$ can be reduced for a sufficiently large $\beta_\calC$ partly because the response of polarization to the isocurvature perturbation is different from that to the adiabatic perturbation. This as a result also helps in reducing the errors in $A_\calR$ by breaking the $\tau$-$A_\calR$ degeneracy. We can see that the estimation of some of the $\Lambda$CDM parameters can well be affected by $\calO(10)\%$ in existence of the cross-correlation, and the complete ignorance of the cross-correlation could result in the misinterpretation of the underlying cosmological model.

Before concluding this section, let us mention here that $\beta_\calC$ can have either sign. 
For instance, for the concrete example in the next section, the sign of $\beta_\calC$ can change depending on the initial displacement angle of the axion. We however checked that a negative $\beta_\calC$ did not change our final conclusion: in view of the forthcoming CMB data, $|\beta_\calC|\gtrsim \calO(0.1)$ would be required for the forthcoming CMB experiment to be sensitive to the isocurvature cross-correlation, and the $\Lambda$CDM cosmological parameter estimations can well be affected at the order of $\calO(10)\%$ in existence of the isocurvature cross-correlation power spectrum.

\section{Correlated axion isocurvature perturbation}
\label{example}

Knowing that the forthcoming CMB data can potentially probe the correlated isocurvature perturbation, it is of interest to study a concrete model which indeed leads to the non-trivial cross-correlation power spectrum. 
For an illustration purpose, we discuss the fluctuations of the axion-like field that gravitationally couples to the inflaton field with the coupling constant $g$ to which the the cross-correlation power spectrum amplitude $A_\calC$ turns out to be proportional. $A_\calC$ and $g$ hence share the common features such as the optimally sensitive scale inferred from the behavior of $\partial C_\ell/ \partial \ln A_\calC \propto \partial C_\ell/\partial \ln g$.  We assume the energy density of the axion is negligible during the inflation and its fluctuations can cause the isocurvature perturbation while the dominant curvature perturbation arises from the inflaton fluctuations. 

We consider a Peccei-Quinn (PQ) field \cite{pq2,pec}
\begin{equation}
\phi = \frac{re^{i\theta}}{\sqrt{2}} \, ,
\end{equation}
and the potential $V_\text{axion}(\phi) = \lambda \left( |\phi|^2 - f_a^2/2 \right)^2$, where the axion decay constant $f_a$ represents the PQ symmetry breaking scale. Then the radial field is settled down at the minimum $f_a$, and we identify the axion as the angular field, $a \equiv f_a\theta$. For an illustration purpose, we consider a concrete toy model with the gravitationally induced interaction
\begin{equation}
V_\text{int}(\chi,\phi) = g \frac{\chi\phi^4}{\mpl}+h.c.+c \, ,
\end{equation}
where $\chi$ is the inflaton field and the constant $c$ is introduced to let the potential vanish at the minimum. %Note that such a non-renormalizable coupling induces the gravitationally induced mass characterizing the explicit symmetry breaking scale of order $m_g^2 \sim 8 g f_a^2 \chi_0 /\mpl$, and this is required to be smaller than the Hubble scale during the inflation $m_\theta^2 \ll H^2$ for the axion fluctuations not to be overdamped.
The following discussion does not heavily depend on the exact form of the inflaton potential $V_\text{inf}(\chi)$, except that the vacuum expectation value of the inflaton should vanish after inflation. We assume $\langle \chi \rangle=0$ after inflation so that this non-renormalizable coupling term does not lead to the explicit symmetry breaking axion mass after inflation not to upset the solution to the strong CP problem \cite{pq2,pec,wein,wil,rus,car}. To keep the generality of our analysis, we just assume that the inflaton sector leads to the adiabatic perturbation with desired amplitude and running.

The Lagrangian of our interest reads
\begin{equation}
\calL = \sqrt{-g} \left[ -\frac{1}{2}(\partial_\mu\chi)^2 - |\partial_\mu\phi|^2 - V_\text{inf}(\chi) - V_\text{axion}(\phi) - V_\text{int}(\chi,\phi) \right] \, .
\end{equation}
To find the cross-correlation power spectrum, we expand the Lagrangian including field fluctuations up to second order. A straightforward and clear way to compute the cross-correlation is to associate the inflaton with the curvature perturbation $\calR$ and the axion with the isocurvature perturbation $\calI$. Then the cross-correlation comes from the interaction between them. Thus we can naturally work in the interaction picture where we can compute the cross-correlation from the interaction Hamiltonian \cite{insteve}. For simplicity, we assume that the radial field $r$ is completely settled down at the minimum $f_a$ so that we can concentrate on the inflaton $\chi$ and the the angular fluctuation $\theta$, and their cross-correlation. Then, the Hamiltonian at quadratic order is
\begin{align}
\calH & = \frac{a^3}{2}\dot\chi^2 + \frac{a}{2}(\nabla\chi)^2 + \frac{a^3f_a^2}{2}\dot\theta^2 + \frac{af_a^2}{2}(\nabla\theta)^2 + a^3\delta V_\text{inf}(\chi) - 4a^3g\frac{\chi_0f_a^4}{\mpl}\cos(4\theta_0)\theta^2 - 2a^3g\frac{f_a^4}{\mpl}\sin(4\theta_0)\chi\theta \, ,
\end{align}
where $\delta V_\text{inf}$ denotes the quadratic part of $V_\text{inf}$ and the subscript 0 means the background value. Note that, in $\calH$, the last term corresponds to the interaction Hamiltonian $\calH_I$ and the rest is the free, kinematic part $\calH_0$. Now, promoting the field fluctuations to the operators and decomposing them in terms of their own, independent creation and annihilation operators as
\begin{equation}
\begin{split}
\chi_{\boldsymbol{k}} & = a_{\boldsymbol{k}}u_k + a_{-{\boldsymbol{k}}}^\dag u_k^* \, ,
\\
\theta_{\boldsymbol{k}} & = b_{\boldsymbol{k}}v_k + b_{-{\boldsymbol{k}}}^\dag v_k^* \, ,
\end{split}
\end{equation}
with $a_{\boldsymbol{k}}$ and $b_{\boldsymbol{k}}$ satisfying the standard commutation relations
\begin{equation}
\left[ a_{\boldsymbol{k}}, a_{\boldsymbol{q}}^\dag \right] = \left[ b_{\boldsymbol{k}}, b_{\boldsymbol{q}}^\dag \right] = (2\pi)^3\delta^{(3)}({\boldsymbol{k}}-{\boldsymbol{q}}) \, , \quad \text{otherwise zero} \, .
\end{equation}
Then the mode function equations follow from $\calH_0$ and are given by
\begin{equation}
\begin{split}
u_k'' - \frac{2}{\tau}u_k' + \left( k^2+\frac{m_\chi^2}{H^2\tau^2} \right)u_k & = 0 \, ,
\\
v_k'' - \frac{2}{\tau}v_k' + \left( k^2+\frac{m_\theta^2}{H^2\tau^2} \right)v_k & = 0 \, ,
\end{split}
\end{equation}
where we have used the Fourier modes, $d\tau = dt/a$ is the conformal time, $m_\chi^2 \equiv \partial^2(\delta V_\text{inf})/\partial\chi^2$ and
\begin{equation}\label{axionmass}
m_{\theta}^2 \equiv -8g\frac{\chi_0}{\mpl}f_a^2\cos(4\theta_0) \, .
\end{equation}
The solutions of these equations are well known: the first kind Hankel function solution with the index determined by the mass,
\begin{equation}
\begin{split}
u_k(\tau) & = -i\exp\left[ i\left(\nu_\chi+\frac{1}{2}\right)\frac{\pi}{2} \right] \frac{\sqrt{\pi}}{2}H(-\tau)^{3/2} H_{\nu_\chi}^{(1)}(-k\tau) \, ,
\\
v_k(\tau) & = -\frac{i}{f_a}\exp\left[ i\left(\nu_\theta+\frac{1}{2}\right)\frac{\pi}{2} \right] \frac{\sqrt{\pi}}{2}H(-\tau)^{3/2} H_{\nu_\theta}^{(1)}(-k\tau) \, ,
\end{split}
\end{equation}
where $\nu_{\chi \, (\theta)} \equiv \sqrt{9/4-m_{\chi \, (\theta)}^2/H^2}$.

Now it is straightforward to compute the cross-correlation $\left\langle \chi_{{\boldsymbol{k}}_1}(t)\theta_{{\boldsymbol{k}}_2}(t) \right\rangle$ using the interaction Hamiltonian $\calH_I$
\begin{align}
\left\langle \chi_{{\boldsymbol{k}}_1}(t)\theta_{{\boldsymbol{k}}_2}(t) \right\rangle & = (2\pi)^3 \delta^{(3)}({\boldsymbol{k}}_1+{\boldsymbol{k}}_2) \frac{2\pi^2}{k^3} \calP_{\chi\theta}
\nonumber\\
& = \Re \left[ 4ig\frac{f_a^4}{\mpl}\sin(4\theta_0) \int_{t_\text{in}}^t dt' a^3(t') \int \frac{d^3q}{(2\pi)^3} \left\langle 0 \left| \chi_{{\boldsymbol{k}}_1}(t)\theta_{{\boldsymbol{k}}_2}(t) \chi_{\boldsymbol{q}}(t')\theta_{-{\boldsymbol{q}}}(t') \right| 0 \right\rangle \right] \, .
\end{align}

Further, defining the curvature and isocurvature perturbations as
\begin{align}
\calR & = -\frac{H}{\dot\chi_0}\chi \, ,
\\
\calI & = 2\frac{\Omega_a}{\Omega_m} \frac{\theta}{\theta_0} \, ,
\end{align}
we can find
\begin{equation}
\calP_{\calC} = -g\pi\sin(4\theta_0) \frac{f_a^3}{\mpl H^2} \Re \left[ i\int_0^\infty \frac{dx}{x} H_{\nu_\chi}^{(2)}(x) H_{\nu_\theta}^{(2)}(x) \right] \sqrt{\calP_\calR\calP_\calI} \, ,
\end{equation}
The factor $\Omega_a/\Omega_m$ arises because we are here interested in the isocurvature perturbation between the radiation and the non-relativistic matter, and the non-adiabatic fluctuations arise from the fluctuations of axion which contributes to the matter density with a fraction $\Omega_a/\Omega_m$. $\Omega_a/\Omega_m$ depends not only on the axion parameters but also on the entropy dilution factor and we just treat it here as a free parameter. The numerical factor involving the integration of Hankel functions is $\sim -0.45$\footnote{One can evaluate this integral, at the leading order, at an arbitrary value of $x$ as long as $e^{-1/\xi} < x < 1$ where $\xi$ is the typical size of the slow-roll parameter \cite{gongsp,ks03}.}.
Note that the $k$-dependence of $\calP_{\calC}$ is just that of $\sqrt{\calP_\calR\calP_\calI}$. This however is not generic but merely due to our particular choice of the coupling, simply $\chi\theta$ without any derivatives. We take only the leading order term $\chi\theta$, and the $k$-dependence of the cross-correlation power spectrum correspondingly becomes same as that of $\sqrt{\calP_\calR\calP_\calI}$. If there are any derivative operators acting on $\chi$ and/or $\theta$, however, we have additional $k$-dependence induced from those derivatives and another form of the $k$-dependence of $\calP_{\calC}$ shows up.
We saw in the last section that $|\beta_\calC|\gtrsim \calO(0.1)$ would be desired for the forthcoming CMB data to be able to probe the correlated isocurvature perturbation. For our toy model under discussion, taking both $f_a$ and $\chi_0$ to be $\calO(\mpl)$ for a simple estimate, we can see that  $H \lesssim \mpl $ leads to $|\beta_\calC|\gtrsim \calO(0.1)$ with $g \sin(4\theta_0)\sim \calO(0.1)$. Note that such a non-renormalizable coupling induces the gravitationally induced mass characterizing the explicit symmetry breaking scale of order $m_{\theta}^2 \sim 8 g f_a^2 \chi_0 /\mpl$, and $ |\beta_\calC| < |\sin (4 \theta_0)| /8$ is required for $m_\theta^2\ll H^2$ not to overdamp the axion fluctuations.

%Note, on the other hand, there is a bound $ |\beta_\calC| < |\sin (4 \theta_0)| /8$ required for $m_\theta^2\ll H^2$. 
%$\beta \sim g \sin 4 \theta (f_a/M_p)^3 (M_p/H)^2$ can be large enough for the Planck-like satellite to be capable of constraining the cross-correlation power specturm. For instance, taking $f_a,\chi$ to be of oder the Planck scale for a simple estimate,

%% \kenji{check: we infer that $\beta$ would not exceed $\calO(10^{-4})$ while satisfying $m^2\ll H_{inf}^2$} unless we can take the inflaton amplitude less than the Planck scale which in principle possible. We however saw in the last section (see for instance Table \ref{normerror}) that, even for $\beta \lesssim 10^{-4}$, the CMB can be sensitive to the existence of the correlated isocurvature perturbations and complete ignorance of the cross-correlation power can affect the parameter estimations by the order of $\calO(10)\%$  

Before concluding our discussions on the axion-inflaton coupling and the resulting cross-correlation power spectrum, let us briefly discuss more general cases. It is straightforward to extend our discussions to other forms of the gravitationally induced couplings between the inflaton and PGB by applying similar steps outlined in our concrete example. For instance, for the PQ-like field $\phi=re^{i\theta}/\sqrt{2}$ with the global symmetry spontaneously broken at the scale $f_a$, the gravitationally induced coupling between the inflaton $\chi$ and $\phi$ of the form 
\begin{equation}
g \frac{\chi^m\phi^n}{\mpl^{m+n-4}}
\end{equation}
leads to the cross-correlation factor $\beta_\calC$
\begin{equation}
\beta_\calC \sim  g \sin \left(n\theta_0\right)  \left(\frac{\chi_0}{\mpl}\right)^{m-1} \left(\frac{f_a}{\mpl}\right)^{n-1} \left(\frac{\mpl}{H}\right)^2 \, .   
\end{equation}
The gravitationally induced mass during the inflation is 
\begin{equation}
m_{\theta}^2 \sim g \frac{n^2}{2^{n/2-1}} \frac{\chi_0^m f_a^{n-2}}{\mpl^{m+n-4}} \, .
\end{equation}
$m_{\theta}^2\ll H^2$ is required for the axion-like field fluctuations not to be suppressed during the inflation, which gives a bound on $\beta_\calC$
\begin{equation}
\beta_\calC \ll \sin (n\theta_0) \frac{2^{n/2-1}}{n^2} \frac{f_a}{\chi_0} \, .
\end{equation}
While the exact values of $f_a$ and $\chi_0$ would be heavily model dependent, we can see that, for an observable isocurvature cross-correlation power spectrum, the symmetry breaking scale $f_a$ would be preferred to be close to the Planck scale \cite{kiwoon,banks,sv}. 
\\
\\
%\section{Conclusions}
We have studied the potential power of the forthcoming CMB data on the isocurvature cross-correlation with the emphasis on the CMB polarization data. Motivated by its possibility to probe the cross-correlation power spectrum, we have derived the analytic form of the cross-correlation power spectrum in a concrete axion-like model. We have explicitly calculated the amplitude and scale dependence and obtained the upper bound for the cross-correlation power amplitude in terms of the axion parameters.
Our study presented here is not meant to be only for the axion but can be extended more generally to any axion-like fields in a straightforward manner which can be possibly abundant in the early Universe. Further explorations of such light degrees of freedom and cosmological probes for them would deserve further studies. The issue on the particle theoretical model building of the axion coupling along with its consequences in view of the forthcoming cosmological (CMB and large scale structure) observables will be discussed elsewhere.

\subsection*{Acknowledgement}
We thank Jai-chan Hwang, Kwang Sik Jeong and Eiichiro Komatsu for the useful discussions, and the Munich Institute for Astro- and Particle Physics of the DFG cluster of excellence ``Origin and Structure of the Universe'', Korea Astronomy and Space Science Institute and Institute for Basic Science for hospitality while this work was carried out. 
\\
This work was supported in part by Institute for Basic Science (IBS-R018-D1(KK)), Starting Grant through the Basic Science Research Program of the National Research Foundation of Korea (2013R1A1A1006701(JG)) and Grant-in-Aid for Scientific Research (No. 24340048(KI), 24540267(TM)) from the Ministry of Education, Sports, Science and Technology (MEXT) of Japan.
JG also acknowledges the Max-Planck-Gesellschaft, the Korea Ministry of Education, Science and Technology, Gyeongsangbuk-Do and Pohang City for the support of the Independent Junior Research Group at the Asia Pacific Center for Theoretical Physics.

%%%%%%%%%%%%%%%%%%%%%%%%%%%%%%%%%%%%%%%

\end{document}